\documentclass[11pt]{article}
\usepackage{graphicx,amsmath,amsfonts,amssymb,dcolumn}

\begin{document}

\title{\textbf{On the topological reduction from the affine to the orthogonal gauge theory of gravity}}
\author{\textbf{R.~F.~Sobreiro$^1$}\thanks{sobreiro@if.uff.br} \ , \textbf{V.~J.~Vasquez Otoya$^2$}\thanks{vjose@cbpf.br}\\\\
\textit{{\small $^1$UFF $-$ Universidade Federal Fluminense,}}\\
\textit{{\small Instituto de F\'{\i}sica, Campus da Praia Vermelha,}}\\
\textit{{\small Avenida General Milton Tavares de Souza s/n, 24210-346 Gragoat\'a,}}\\
\textit{{\small Niter\'oi, Brasil.}}\and
\textit{{\small $^2$CBPF, Centro Brasileiro de Pesquisas F\'{\i}sicas,}} \\
\textit{{\small Rua Xavier Sigaud 150, 22290-180 Urca,}}\\
\textit{{\small Rio de Janeiro, Brasil}}}
\date{}
\maketitle

\begin{abstract}
\noindent Making use of the fibre bundle theory to describe metric-affine gauge theories of gravity we are able to show that metric-affine gauge theory can be reduced to the Riemann-Cartan one. The price we pay for simplifying the geometry is the presence of matter fields associated with the nonmetric degrees of freedom of the original setup. Also, a possible framework for the construction of a quantum gravity theory is developed along the text.
\end{abstract}


\section{Introduction}\label{intro}

Gravity under the framework of gauge theories, first emerged with the seminal works of R.~Utiyama and T.~W.~B.~Kibble \cite{Utiyama:1956sy,Kibble:1961ba}, is of great interest mainly because it puts gravity almost at the same level as the other fundamental interactions. A curious difference between gauge gravity and the other interactions is that, instead of a unitary group, gravity is based on the Lorentz non-compact group $SO(1,3)$ or the special orthogonal group $SO(4)$, which enables a direct relation between spacetime and gauge space. This identification is performed by the vierbein field which is directly related to the spacetime metric tensor. The gauge field is the so called spin-connection which is associated, together with the vierbein, to the spacetime connection. Through the spin-connection, spinor fields can now be minimally coupled to the Einstein-Hilbert action and work as a source for the Cartan torsion.\\\\
It is a fact that the Einstein-Hilbert action, in the first and second order formalisms, is not a renormalizable action \cite{'tHooft:1974bx,Deser:1974cz,Deser:1974cy}. To face the renormalizability problem and define a consistent quantum field theory for gravity there were proposed several alternatives by extending the Einstein-Hilbert action to more general ones \cite{Lovelock:1971yv,Stelle:1976gc,Mardones:1990qc,Zanelli:2005sa}. However, their quantum consistency is yet a mistery. Another generalization, that transcends the action level, is to consider more general groups for the gauge symmetry that also allows the identification of the gauge space with the spacetime cotangent bundle, in particular we recall the so called metric-affine gravities, based on the affine gauge group $A(d,\mathbb{R})$, originated with F.~W.~Hehl and colaborators \cite{Hehl:1976my} and extensively developed by several authors, see for instance \cite{Hennig:1981id,Tresguerres:1995un,Obukhov:1996pf,Tapia:1998as,King:2000ha,Tresguerres:2000qn,Obukhov:2002tm}, to name a few of the most relevant works. See also \cite{Hehl:1994ue} and references therein for a very complete review on metric-affine gravities. This class of theories also requires extra terms, apart from the Einstein-Hilbert one, to account the quantum divergences and extra degrees of freedom. The present article is in fact devoted to the study of some geometrical properties of this class of theories and the development of a possible scenario for its quantization.\\\\
We consider a general framework for metric-affine gravities obtaining results that are independent of the specific dynamical equations. In fact, they depend exclusively on the topological features of a gauge theory and its description in terms of fibre bundles \cite{Kobayashi,Nakahara:1990th,Singer:1978dk,CottaRamusino:1985ad,Falqui:1985iu,Daniel:1979ez,Bertlmann:1996xk}. Specifically, we construct the usual principal bundles for a gauge theory in a generic real differential manifold \cite{Nakahara:1990th,Singer:1978dk,Daniel:1979ez,Bertlmann:1996xk} with structure group being the affine group. The \emph{vielbein} one-form field and that what would be the metric in gauge space, called \emph{premetric} zero-form, are both introduced as matter fields not related with spacetime, \emph{i.e.} vielbeins are not identified with \emph{coframes} in the cotangent bundle of the spacetime manifold and the premetric is not associated with the spacetime metric tensor, called here simply by \emph{metric}. In fact, to avoid the characterization of the premetric with the metric we define also a dual gauge space, allowing the construction of gauge invariant quantities with no reference to the premetric as a geometrical quantity. Thus, vielbein and premetric are both genuine matter fields. The vielbein-coframe identification as well as the premetric-metric one can be performed afterwards and the above referred bundles are identified with the usual coaffine bundle \cite{Hennig:1981id,Trautman:1979cq,Trautman:1981fd} and its respective dynamical connection bundle. The last one is commonly studied in Yang-Mills theories \cite{Singer:1978dk,CottaRamusino:1985ad,Falqui:1985iu,Bertlmann:1996xk} but, to our knowledge, not for gravity specifically. It turns out that this scenario is very suitable for the construction of a quantum gauge theory of metric-affine gravity where the quantum sector is characterized by a gauge theory in a curved spacetime in which the gauge space and the cotangent bundle are not related to each other while the classical sector is characterized by such identification. Thus, after the identification, gravity as a dynamical geometric theory arises, as it is expected from gravity. It is also expected that both sectors do not overlap, allowing then the compatibility of the principles of general relativity with those of quantum field theory, see for instance \cite{Sobreiro:2007pn}. However, we emphasize that the mechanism triggering the identification still lacks and the present work will not deal with such issues.\\\\
The main result of the present work concerns the geometric and topological properties of the above discussed framework. For that we evoke the theory of fibre bundles for the description of gauge theories. The usual construction of a principal bundle for gauge theories takes the spacetime manifold as the base space and the structure group to be the gauge symmetry group \cite{Nakahara:1990th,Daniel:1979ez}. This bundle gives rise to the gauge connection one-form. However, to give dynamics to the gauge connection another principal bundle has to be considered where the base space is now the space of all independent gauge connections and the fibre is represented by the gauge orbit \cite{Nakahara:1990th,Singer:1978dk,Bertlmann:1996xk}. The total space is then the space of all possible gauge connections that can be defined on the original principal bundle. In the present case the first principal bundle is called by \emph{affine gauge bundle} while the last by \emph{dynamical affine gauge bundle}. In both cases, the gauge group space is regarded as independent of the cotangent bundle of the spacetime manifold. Thus, the prescription is that of a gauge theory on a curved spacetime. Now, in those bundles, one can define two tensor representations, the above discussed vielbein and premetric fields, that afterwards will be identified with coframes and metric, respectively. Thus, the affine gauge bundle is mapped into the usual coaffine bundle \cite{Trautman:1979cq,Trautman:1981fd,Hennig:1981id} and the dynamical affine gauge bundle is mapped the correspondent one constructed from the coaffine bundle. Gravity then emerges under the scope of metric-affine geometry \cite{Hehl:1976my,Hehl:1994ue} where nonmetric degrees of freedom are important sectors of the theory. In despite of such identification, some well known theorems concerning the contraction of principal bundles can be applied for both gauge bundles. The affine group can then be contracted down to the orthogonal one, resulting on simpler bundles and a simpler gauge connection. The main properties allowing such reduction are the non-compact nature of the affine group and the symmetric character of the coset space $A(d,\mathbb{R})/O(d)$. The resulting gauge theory is an orthogonal one that, if one assumes the gauge space-cotangent bundle identification, can be mapped into the usual Riemann-Cartan gravity. However, the nonmetric degrees of freedom belonging to the original setup (the translational and symmetric sectors of the full affine gauge connection) do not simply drop, specially in the dynamical bundle, they survive as extra tensor representations, \emph{i.e.} additional matter fields. Thus, we prove that the metric-affine gauge theory can be naturally reduced to a Riemann-Cartan one with new matter fields.\\\\
We remark that, for the coframe bundle case, the reduction to the orthogonal bundle is already known in the current literature \cite{Kobayashi,Nash:1983cq,McInnes:1984kz}. The unique difference here is the independence between the gauge space and the cotangent bundle. We recall that the present results are a formal generalization of those heuristically discussed in \cite{Sobreiro:2007cm} for a particular case of metric-affine gravities. Also, there are several methods where spontaneous symmetry breaking has been employed to provide the reduction of the affine bundle to smaller ones in which the vielbein arise from the original connection \cite{Hennig:1981id,Trautman:1979cq,Trautman:1981fd}. For the dynamical connection bundle, on the other hand, the results concerning its reduction are original as well as the respective physical interpretations.\\\\
It is worth mention that alternatives to describe gravity differently from the usual coframe or even coaffine bundles have been made. In particular, the framework developed in \cite{Lord:1986xi,Lord:1986at,Lord:1988nd} is of great interest due to the fact that every gravity field appearing in the first order formalism is contained in the geometrical setting of the fundamental principal bundle of the model. Basically, connection and vielbein are both obtained from the coframes of the total space and then all gravity dynamical variables arise from geometrical definitions. Actually, all gauge theories of gravity based on fibre bundle construction use to assume that gauge space and spacetime are identified. Always the (spacetime) tangent bundle is the structure where gauge connection and vielbein are defined. Even if connection and vielbein are unified in a single object like a generalized connection \cite{Hennig:1981id,Trautman:1979cq,Trautman:1981fd} or like more sophisticated bundles \cite{Lord:1986xi,Lord:1986at,Lord:1988nd} where the connection is a particular type of ordered basis on the total space.\\\\
We have already commented on the difference between the present work and those of \cite{Hennig:1981id,Trautman:1979cq,Trautman:1981fd}. Concerning the approach of \cite{Lord:1986xi,Lord:1986at,Lord:1988nd} it differs from ours in, basically, three fundamental points. First, the base space is the spacetime defined as the coset space of the total space that encodes diffeomorphisms and gauge symmetries and thus gauge and spacetime aspects are unified in a single symmetry group. Notwithstanding, in the present article we split spacetime and gauge spaces in order to avoid the mix of quantum field variables and spacetime coordinates (at quantum level). In fact, the framework developed in \cite{Lord:1986xi,Lord:1986at,Lord:1988nd} might also have to deal with the problem of open algebras if one attempts to quantize the model \cite{Henneaux:1985kr}. Second, it is nowadays established that a quantum gauge theory should be based on the dynamical bundle, where the space of independent connections should be the base space \cite{Nakahara:1990th,Singer:1978dk,CottaRamusino:1985ad,Falqui:1985iu,Daniel:1979ez,Bertlmann:1996xk} and thus not the spacetime. Moreover, as a third argumentation, the works \cite{Lord:1986xi,Lord:1986at,Lord:1988nd} have as motivation the unification of spacetime and gauge symmetries as well as the union of connection and vielbein structures. Our aim is actually not to unify them but separate them from the spacetime structure. Only at classical level those structures might be identified through dynamical effects.\\\\ 
This work is organized as follows: In Sect.\ref{MAG} the basic properties of metric-affine gauge theories of gravity are developed. In this section we fix the notation, conventions and also provide our interpretation and subtle modifications of the theory. In Sect.\ref{FBT} we summarize the fibre bundle description of gauge theories in curved spaces. Also in this section we state three theorems concerning fibre bundles that will be crucial to our results. Sect.\ref{FBMAG} is devoted for the construction of the relevant principal bundles for the theory described in Sect.\ref{MAG} and specify the main difference of a generic gauge theory and a gauge theory for gravity. Then, the fibre bundle theorems of Sect.\ref{FBT} are applied in order to contract the metric-affine gauge gravities resulting in the usual Riemann-Cartan one with extra matter fields associated with the original nonmetric degrees of freedom. Finally, in Sect.\ref{FINAL} we display our conclusions and further discussions.


\section{Metric-affine gauge theory of gravity}\label{MAG}

In this section we provide the basic properties of the metric-affine gauge theory of gravity \cite{Hehl:1976my,Hehl:1994ue}. In particular, we discuss our conventions, definitions and further important aspects that will be used in the following sections. We remark that some of our considerations are slightly different from the standard literature on gauge theories of gravity \cite{Kibble:1961ba,Mardones:1990qc,Zanelli:2005sa,Hehl:1976my,Hehl:1994ue}.

\subsection{Group structure}\label{grstr}

The affine group is defined as
\begin{equation}
A(d,\mathbb{R})\cong
GL(d,\mathbb{R})\ltimes\mathbb{R}^d\;,\label{affine0}
\end{equation}
where $GL(d,\mathbb{R})$ is the general linear group of $d\times d$
invertible real matrices and $\mathbb{R}^d$ stands for the group of
translations. Generically, we have for the Lie algebra decomposition $a=gl\oplus r$,
\begin{eqnarray}
\left[gl,gl\right]&\subseteq&gl\;,\nonumber\\
\left[r,r\right]&\subseteq&\emptyset\;,\nonumber\\
\left[r,gl\right]&\subseteq&r\;,\label{alg000}
\end{eqnarray}
where $gl$ is the algebra of the $GL(d,\mathbb{R})$ and $r$ the
algebra of the $\mathbb{R}^d$ translations. We remark that the affine group is not a semi-simple group due
to the translational sector. Further, the algebra decomposition
\eqref{alg000}, implies that the coset space $\mathbb{R}^d\cong
A(d,\mathbb{R})/GL(d,\mathbb{R})$ is a symmetric space and $GL(d,\mathbb{R})$ is a noncompact stability group of $A(d,\mathbb{R})$. Also, it is trivial to see that the translational sector has $d$ dimensions while the general linear group has $d^2$ dimensions and thus, the affine group is provided with $d(d+1)$ dimensions.\\\\
The general linear group can also be decomposed as
\begin{equation}
GL(d,\mathbb{R})\cong K(d)\otimes O(d)\;,\label{cong1}
\end{equation}
where $K(d)$ is the space defined from the symmetric part of the
algebra of the general linear group and possesses $d(d+1)/2$
dimensions. Formally, $K(d)$ is the coset space $K(d)\cong GL(d,\mathbb{R})/O(d)$
and, differently of $\mathbb{R}^d$, it does not form a group. The
orthogonal group $O(d)$, with $d(d-1)/2$ dimensions, is a
maximal compact subgroup of the affine group as well as for the
general linear group. The algebra decomposition can be described
through
\begin{eqnarray}
\left[o,o\right]&\subseteq&o\;,\nonumber\\
\left[k,k\right]&\subseteq&o\;,\nonumber\\
\left[k,o\right]&\subseteq&k\;,\label{coset0}
\end{eqnarray}
where $o$ stands for the algebra of the orthogonal $O(d)$ group and
$k$ for the algebra of the coset space $K(d)$. From
(\ref{coset0}) one can infer that $K(d)$ forms a
symmetric space.\\\\
The elements of the affine group are here represented by exponentials according to
\begin{equation}
\mathcal{U}=\mathrm{e}^{\zeta(x)}=\mathrm{e}^{\alpha^a_{\phantom{a}b}(x)T^b_{\phantom{b}a}\oplus\xi^a(x)P_a}\;.\label{group1}
\end{equation}
where we are considering the algebraic notation of the adjoint representation of the affine group.  In fact, in this article, we are exclusively working in the adjoint representation. Thus, small Latin indices are vector indices of the $d$-dimensional differential manifold defined by the affine group, $a\in\{1,2,\ldots,d\}$. In \eqref{group1} the parameter $\zeta$ is a $A(d,\mathbb{R})$ algebra-valued quantity $\zeta=\zeta^AS_A$ where $S_A$ represents the group generators in adjoint representation with $A\in\{1,2,\ldots,d(d+1)\}$. At the same level, $\alpha^a_{\phantom{a}b}$ and $\xi^a$ are algebra-valued parameters associated with the group element $U\in GL(d,\mathbb{R})$ and $u\in\mathbb{R}^d$, respectively, while $T^b_{\phantom{b}a}$ and $P_a$ the respective generators. Further, it is a general property of Lie groups that the element $\mathcal{U}$ can always be decomposed as $\mathcal{U}=uU$. Furthermore, under decomposition \eqref{coset0} the elements $U$ can be splitted according to $U=LC$, with $L=\mathrm{e}^{h^a_{\phantom{a}b}(x)\Sigma^b_{\phantom{b}a}}\in O(d)$ and $C=\mathrm{e}^{b^a_{\phantom{a}b}(x)\lambda^b_{\phantom{b}a}}\in K(d)$, where the notation speaks for itself.\\\\
It is important to understand that the translational sector is usually defined, according to the semidirect product in \eqref{affine0}, as an orthogonal subgroup of the affine group due to its action on spacetime coordinates, $x^\prime=Ux+\xi$, with $\xi$ finite or infinitesimal. In this case the M\"obius representation \cite{Hehl:1994ue,KobayashiX} is very usefull. However, since we are constructing a gauge theory, \emph{i.e.}, we shall not deal with spacetime transformations, we rather use the exponential representation above described to act over fields in the cotangent space. It turns out that the semidirect product in \eqref{affine0} would only take place for vector representations on the tangent space. In practice, we can think of a direct product between the general linear and the translational groups. Let us take then a full affine field $\Psi=\psi\oplus\phi$ such that $\psi=\psi^a_{\phantom{a}b}T^b_{\phantom{b}a}$ and $\phi=\phi^aP_a$. Its group action can be written as
\begin{equation}
\Psi\longmapsto\mathcal{U}^{-1}\Psi\;\mathcal{U}\;,\label{field0}
\end{equation}
However, its decomposition is obtained by requiring invariance of the translational sector through the right action of the group,
\begin{equation}
\Psi\longmapsto\mathcal{U}^{-1}\psi\;\mathcal{U}\oplus\mathcal{U}^{-1}\phi\;=\;\mathcal{U}^{-1}\psi\;\mathcal{U}\oplus U^{-1}\phi\;.\label{field1}
\end{equation}
The decomposition rule \eqref{field1} ensures the consistency of the condensed notation \eqref{field0} which shall always be used in this article.\\\\
For consistency let us also write the relations \eqref{alg000} in terms of the generators of the affine group,
\begin{eqnarray}
\left[T^b_{\phantom{b}a},T^d_{\phantom{d}c}\right]&=&\left(\delta_a^d\delta^b_e\delta^f_c-\delta^f_a\delta^b_c\delta^d_e\right)T^e_{\phantom{e}f}\;,\nonumber\\
\left[P_a,P_b\right]&=&0\;,\nonumber\\
\left[T^b_{\phantom{b}a},P_c\right]&=&-\delta^b_cP_a\;,\label{alg000x}
\end{eqnarray}
while the correspondent relation for \eqref{coset0} follows from \eqref{alg000x} by suitable (anti-)symmetrization of the indices and shall not be explicitly required in what follows.

\subsection{Gauge structure}

Our scenario is a  $A(d,\mathbb{R})$ gauge theory over a $d$-dimensional differential manifold
$\mathbb{M}^d$. The internal space defined by the gauge group is, at first, independent of the space-time \cite{Mardones:1990qc,Zanelli:2005sa,Hehl:1994ue}. Thus, the tangent bundle defined in $\mathbb{M}^d$ is not yet identified with the internal gauge space.

\subsubsection{Gauge connections, covariant derivatives and field strengths}\label{gauge00}

The main ingredient of a gauge theory is the gauge potential
one-form. For the affine gauge theory one may define
\begin{equation}
Y=Y^AS_A=\omega^a_{\phantom{a}b}T^b_{\phantom{b}a}\oplus E^aP_a\;,\label{conn01}
\end{equation}
where $\omega^a_{\phantom{a}b}$ is the $GL(d,\mathbb{R})$ connection
and $E^a$ the translational one. It is important to keep in mind
that $E^a$ and the vielbein field are essentially different. They can only be
identified if the translation local symmetry is broken to a global
one or a compensating field is introduced, see for instance
\cite{Hehl:1994ue,Stelle:1979aj}. We shall return to this point still on Sect.\ref{Ee}.\\\\
The connection $Y$ transforms under local gauge transformation in
the standard manner,
\begin{equation}
Y\longmapsto\mathcal{U}^{-1}\left(\mathrm{d}+Y\right)\mathcal{U}\;,\label{gauge02}
\end{equation}
where $\mathrm{d}$ is the space-time exterior derivative. At infinitesimal level the transformation \eqref{gauge02} reduces to
\begin{eqnarray}
Y\longmapsto Y+\nabla\zeta\;,\label{gauge02inf}
\end{eqnarray}
where the covariant derivative is defined as
\begin{eqnarray}
\nabla=\mathrm{d}+Y\;,\label{covder01}
\end{eqnarray}
as long as we are considering the adjoint representation. Transformation \eqref{gauge02inf} decomposes as
\begin{eqnarray}
\omega&\longmapsto&\omega+\mathcal{D}\alpha\;,\nonumber\\
E&\longmapsto&E+\mathcal{D}\xi+E\alpha\;,\label{gauge02infa}
\end{eqnarray}
where the covariant derivative $\mathcal{D}$ is taken only with respect to the $GL(d,\mathbb{R})$ connection
\begin{equation}
\mathcal{D}=\mathrm{d}+\omega\;.\label{covder02}
\end{equation}
In components expression \eqref{gauge02infa} reads
\begin{eqnarray}
\omega^a_{\phantom{a}b}&\longmapsto&\omega^a_{\phantom{a}b}+\mathcal{D}\alpha^a_{\phantom{a}b}\;=\;\omega^a_{\phantom{a}b}+\mathrm{d}\alpha^a_{\phantom{a}b}-\omega^a_{\phantom{a}c}\alpha^c_{\phantom{c}b}+\omega^c_{\phantom{c}b}\alpha^a_{\phantom{a}c}\;,\nonumber\\
E^a&\longmapsto&E^a+\mathcal{D}\xi^a+E^b\alpha^a_{\phantom{a}b}\;=\;E^a+\mathrm{d}\xi^a-\omega^a_{\phantom{a}b}\xi^b+E^b\alpha^a_{\phantom{a}b}\;.\label{gauge02infb}
\end{eqnarray}\\\\
From the definition \eqref{covder01} one can easily compute the field strength two-form
\begin{equation}
\nabla^2=\Theta=\Omega^a_{\phantom{a}b}T^b_{\phantom{b}a}\oplus\Xi^aP_a\;,\label{curv01}
\end{equation}
where $\Omega^a_{\phantom{a}b}$ is the $GL(d,\mathbb{R})$ curvature
while $\Xi^a$ the translational one,
\begin{eqnarray}
\Omega^a_{\phantom{a}b}&=&\mathrm{d}\omega^a_{\phantom{a}b}-\omega_{\phantom{a}c}^a\omega^c_{\phantom{c}b}\;,\nonumber\\
\Xi^a&=&\mathrm{d}E^a-\omega_{\phantom{a}b}^aE^b\;=\;\nabla E^a\;=\;\mathcal{D}E^a\;.\label{curv02}
\end{eqnarray}
In the existing literature the translational sector of the curvature is sometimes also called torsion. However, since in the present construction the vielbein is not directly related to the translational connection, we avoid such classification and call it simply by \emph{translational curvature}.\\\\
The field strength \eqref{curv01} is a covariant quantity obeying a transformation of the type \eqref{field0}. At infinitesimal level it decomposes as
\begin{eqnarray}
\Omega&\longmapsto&\Omega+\Omega\alpha\;,\nonumber\\
\Xi&\longmapsto&\Xi+\Xi\alpha+\Omega\xi\;.\label{gauge03}
\end{eqnarray}
Thus, $\Omega$ is \emph{blind} under the translational sector while $\Xi$ \emph{sees} both sectors of the affine group. Further, under a pure $GL(d,\mathbb{R})$ gauge transformation, both, $\Omega$ and $\Xi$, transform covariantly. However, under a pure translational
transformation, $\Omega$ is invariant while $\Xi$ is not even covariant. This is indeed a direct reflex of the non-semi-simplicity of the translational sector. In fact, this property is associated with the fact that there is no invariant Killing metric for a non-semisimple group, resulting on the impossibility to construct gauge invariant local actions accounting for the translational sector \cite{Hennig:1981id,Aldrovandi:1985gt,Aldrovandi:1988rz}.

\subsubsection{Vielbein as matter and torsion}\label{Ee}

An extra ingredient is always introduced in gravity gauge theories, a translational algebra-valued one-form, the vielbein field $e=e^aP_a$. Under a gauge transformation, it is imposed a matter field type transformation
\begin{equation}
e\longmapsto \mathcal{U}^{-1}e=U^{-1}e\;.\label{gauge01}
\end{equation}
For an infinitesimal transformation expression \eqref{gauge01} reduces to
\begin{equation}
e\longmapsto e-\alpha e\;,\label{gauge01inf}
\end{equation}
that in components reads
\begin{equation}
e^a\longmapsto e^a-\alpha_{\phantom{a}b}^ae^b\;.\label{gauge01infa}
\end{equation}\\\\
The minimal coupling of the vielbein gives rise to what will be the Cartan torsion\footnote{Strictly speaking, $T^a$ is not yet the Cartan torsion since $e^a$ is not yet identified with coframes in spacetime.} \cite{Hehl:1994ue,Cartan}, here generalized to the affine case,
\begin{equation}
T^a=\nabla e^a=\mathcal{D}e^a=\mathrm{d}e^a-\omega^a_{\phantom{a}b}e^b\;.\label{tor01}
\end{equation}\\\\
It is worth mention that a relation between $\Xi^a$ and $T^a$ can be enforced by introducing a zero-form field $\varphi^a$, identifying $E^a$ and $e^a$ \cite{Hehl:1994ue,Trautman:1979cq,Hennig:1981id},
\begin{equation}
e^a=E^a+\nabla\varphi^a=E^a+\mathcal{D}\varphi^a\;,\label{id01}
\end{equation}
where $\varphi^a$ is supposed to transform as an affine vector in group space
\begin{equation}
\varphi\longmapsto U^{-1}\left(\varphi-\xi\right)\approx\varphi-\xi-\alpha\varphi\;.\label{gauge04}
\end{equation}
From \eqref{id01} we have then
\begin{equation}
T^a=\Xi^a+\Omega^a_{\phantom{a}b}\varphi^b\;.\label{id02}
\end{equation}
The introduction of the scalar field $\varphi^a$ allows one to enforce a kind of symmetry breaking of the translational sector \cite{Hehl:1994ue,Stelle:1979aj} through the constraint $\mathcal{D}\varphi^a=0$
and end up with the equality $E^a=e^a$. As announced before, we shall follow another direction and eliminate the translational local symmetry by topological arguments, see Sect.\ref{FBMAG}.\\\\
We emphasize that the name vielbein is used here when referring to the matter field $e^a$ that is not related to coframes. If the gauge symmetry is identified with spacetime then we shall refer to $e^a$ as coframes.

\subsubsection{Metric tensor as matter and nonmetricity}

The usual Einstein-Cartan theory of gravity \cite{Utiyama:1956sy,Kibble:1961ba,Mardones:1990qc,Zanelli:2005sa} is constructed with the local Lorentz group. Obviously, the space defined by this group is of a Minkowski type and thus characterized by a flat metric. On the other hand, in an affine gauge theory, the group space is no longer flat\footnote{In Einstein-Cartan theory $g_{ab}$ is fixed as $\mathrm{diag}(1,-1,-1,-1,\ldots)$. In metric-affine gauge theory even if we start from an orthonormal basis, an affine gauge transformation would always deform such coframe. Thus, the premetric has to be considered as a dynamical field.} and an extra field is required, the premetric symmetric field $g^{ab}$, a zero-form double algebra-valued on the translational sector $g=g^{ab}P_aP_b$. It is crucial to consider that, just like the vielbein, $g^{ab}$ is also independent from any other field. Thus, we take $g^{ab}$ to be a matter field not yet identified with geometry.  The transformation of $g$ is then of matter type,
\begin{equation}
g\longmapsto\mathcal{U}^{-1}g\left(\mathcal{U}^{-1}\right)^T=U^{-1}g\left(U^{-1}\right)^T\;,\label{gtransf}
\end{equation}
which reduces, for infinitesimal transformations, to
\begin{equation}
g^{ab}\longmapsto g^{ab}-\alpha^a_{\phantom{a}c}g^{bc}-\alpha^b_{\phantom{b}c}g^{ac}\;.\label{gtransfinf}
\end{equation}
The minimal coupling of the premetric is
\begin{equation}
Q^{ab}=\nabla g^{ab}=\mathcal{D}g^{ab}=\mathrm{d}g^{ab}-\omega^a_{\phantom{a}c}g^{cb}-\omega^b_{\phantom{b}c}g^{ca}\;,\label{nm}
\end{equation}
and is associated with nonmetricity \cite{Hehl:1976my,Hehl:1994ue} when $g^{ab}$ is identified with the spacetime metric. In the particular case of vanishing nonmetricity the connections are said to be compatible with the metric. Obviously, in general metric-affine gravities the connection is not compatible with the metric.\\\\
Just like the vielbein, we recall that, when we call $g^{ab}$ by premetric we are referring to the matter field that is not related to the metric tensor. If the gauge symmetry is identified with spacetime then we shall refer to $g^{ab}$ simply as metric.

\subsubsection{Dual gauge space and gauge invariants}

Until now, the construction of the metric-affine gravity assumes complete independence between spacetime and gauge space. The theory should then be constructed with the fundamental independent fields $(\omega^a_{\phantom{a}b},E^a,e^a,g^{ab})$. In order to construct gauge invariant quantities one can introduce a dual gauge space represented by a set of dual fields $(E_a,e_a,g_{ab})$ and then avoid the rise and lowering of gauge indices. Since $\omega^a_{\phantom{a}b}$ already lives in both sectors, there is no need to introduce a dual $GL(d,\mathbb{R})$ connection. However, dual group generators are introduced, namely $T_a^{\phantom{a}b}$ and $P^a$, obeying the standard algebra
\begin{eqnarray}
\left[T_a^{\phantom{a}b},T_c^{\phantom{c}d}\right]&=&\left(\delta_a^e\delta^b_c\delta^d_f-\delta^d_a\delta^e_c\delta^b_f\right)T_e^{\phantom{e}f}\;,\nonumber\\
\left[P^a,P^b\right]&=&0\;,\nonumber\\
\left[T_a^{\phantom{a}b},P^c\right]&=&-\delta^c_aP^b\;,\label{alg000xx}
\end{eqnarray}
that, just like representation \eqref{alg000x}, has no reference to the (dual) premetric, avoiding the problems of dealing with open algebras \cite{Henneaux:1985kr}. A dual group element is represented by $\overline{\mathcal{U}}=\exp(\overline{\alpha}\oplus\overline{\xi})=\exp(\alpha^a_{\phantom{a}b}T_a^{\phantom{a}b}\oplus\xi_aP^a)$ while the dual fields, denoted with a bar over them, are expanded as
\begin{eqnarray}
\overline{\omega}&=&-\omega^a_{\phantom{a}b}T_a^{\phantom{a}b};,\nonumber\\
\overline{E}&=&-E_aP^a\;,\nonumber\\
\overline{e}&=&e_aP^a\;,\nonumber\\
\overline{g}&=&g_{ab}P^aP^b\;.\label{dual0}
\end{eqnarray}
where the minus sign in \eqref{dual0} is required for the correct action of the covariant derivative in dual indices, see \eqref{gauge02infb} for the quantity $\alpha^a_{\phantom{a}b}$. Thus, the dual affine gauge connection is defined as $\overline{Y}=-(\overline\omega\oplus\overline{E})$ and its gauge transformation reads
\begin{equation}
\overline{Y}\longmapsto\overline{\mathcal{U}}^{-1}(\mathrm{d}+\overline{Y})\overline{\mathcal{U}}\;,\label{dualconn0}
\end{equation}
resulting in exactly the same infinitesimal gauge transformation \eqref{gauge02infb} apart from sign differences,
\begin{eqnarray}
\omega^a_{\phantom{a}b}&\longmapsto&\omega^a_{\phantom{a}b}-\mathcal{D}\alpha^a_{\phantom{a}b}\;,\nonumber\\
E_a&\longmapsto&E_a-\mathcal{D}\xi_a-E_b\alpha^b_{\phantom{b}a}=E_a-(\mathrm{d}\xi_a+\omega^b_{\phantom{b}a}\xi_b+E_b\alpha^b_{\phantom{b}a})\;.\label{dualconn1}
\end{eqnarray}
As a consequence the covariant derivative, in components, has the same definition as before while for the expanded one a bar has to be employed to account for the dual generators. The curvature generated by the dual approach is decomposed as
\begin{eqnarray}
\overline{\Omega}^a_{\phantom{a}b}&=&-\Omega^a_{\phantom{a}b}\;,\nonumber\\
\Xi_a&=&-(\mathrm{d}E_a+\omega^b_{\phantom{b}a}E_b)\;=\;-\mathcal{D}E_a\;,\label{curvnew}
\end{eqnarray}
where a bar over the dual general linear curvature is necessary due to its ambiguous character. Following the construction of the representation developed in Sect.\ref{grstr}, a dual translational quantity must be invariant under a left action of the group, $\mathcal{U}^{-1}\overline\phi=\overline\phi$. Thus, the dual transformation analogous to \eqref{field1} is
\begin{equation}
\overline{\Psi}\longmapsto\overline{\mathcal{U}}^{-1}\;\overline{\psi}\;\overline{\mathcal{U}}\oplus\overline{\phi}\;\overline{\mathcal{U}}\;=\;\overline{\mathcal{U}}^{-1}\;\overline{\psi}\;\overline{\mathcal{U}}\oplus \overline{\phi}\;\overline{U}\;.\label{field1}
\end{equation}
This is in fact the transformation rule of the dual curvature $\overline{\Theta}=\overline{\Omega}\oplus\overline{\Xi}$.\\\\
The dual matter fields $e_a$ and $g_{ab}$ transform as
\begin{eqnarray}
\overline{e}&\longmapsto&\overline{e}\;\overline{\mathcal{U}}\;=\;\overline{e}\;\overline{U}\;,\nonumber\\
\overline{g}&\longmapsto&\overline{\mathcal{U}}\overline{g}\left(\overline{\mathcal{U}}\right)^T\;=\;\overline{U}\overline{g}\left(\overline{U}\right)^T\;.\nonumber\\
\end{eqnarray}
At infinitesimal level this transformations reduce to
\begin{eqnarray}
e_a&\longmapsto&e_a+\alpha^b_{\phantom{b}a} e_b\;,\nonumber\\
g_{ab}&\longmapsto&g_{ab}+\alpha^c_{\phantom{c}a}g_{cb}+\alpha^c_{\phantom{c}b}g_{ca}\;;\nonumber\\
\end{eqnarray}
the resulting minimal couplings for the dual matter fields are then
\begin{eqnarray}
T_a&=&\mathrm{d}e_a+\omega^b_{\phantom{b}a}e_b\;,\nonumber\\
Q_{ab}&=&\mathrm{d}g_{ab}+\omega^c_{\phantom{c}a}g_{bc}+\omega^c_{\phantom{c}b}g_{ac}\;.\label{dualx}
\end{eqnarray}\\\\
In practice, one can construct gauge invariant quantities by contracting gauge and dual gauge indices. It is not difficult to show that in gauge component notation those quantities are indeed gauge invariant objects even for finite transformations. Careful must be employed for the translational sector due to its non-semisimplicity property \cite{Aldrovandi:1985gt,Aldrovandi:1988rz}. We also remark that the naive definition of low indices as those that transforms with the respective inverse transformation automatically imply on the existence of a metric tensor in gauge space and thus must be avoided. Thus, both sectors have to be distinguished by specific gauge transformations that can be considered inverses only at component level. Standard local gauge invariant actions can then be constructed with suitable combinations of $\omega^a_{\phantom{a}b}$, $e^a$, $e_a$, $g^{ab}$, $g_{ab}$, $\Omega^a_{\phantom{a}b}$, $T^a$, $T_a$, $Q^{ab}$ and $Q_{ab}$.\\\\
The dual gauge space is very important because it avoids the use of the premetric as a metric tensor, as is being our proposition. In fact, a differential manifold can be constructed with no reference to the existence of a metric tensor \cite{Kobayashi,Synge}. Under this situation we can freely consider $g^{ab}$ and $g_{ab}$ as genuine independent matter fields. The same fact holds for $e^a$ and $e_a$, as well as for $E^a$ and $E_a$, as long as the gauge space is not identified with the cotangent bundle of the spacetime. Thus, the duals and the formers are supposed to be independent fields obeying independent dynamical equations.

\subsubsection{Vielbein-coframe identification and geometry}

The idea behind a gravity gauge theory is that, after still unknown dynamical effects, the gauge space is identified space-time \cite{Mardones:1990qc,Zanelli:2005sa,Hehl:1994ue}, and gravity as a geometric dynamical theory emerges. Such identification is obtained by
identifying the vielbein field with coframes in the cotangent bundle associated with the manifold $\mathbb{M}^d$. In fact, at a point $p\in{\mathbb{M}^d}$, a cotangent space $*\mathbb{T}^d_p$ can be defined. The union of all such cotangent spaces along $\mathbb{M}^d$ forms the so called cotangent bundle. Further, in $*\mathbb{T}^d_p$ one can define the collection of all ordered basis (coframes), forming a fibre \cite{Kobayashi,Nakahara:1990th,Nash:1983cq} along $*\mathbb{T}^d_p$. The union of all such fibres forms the coaffine bundle if such coframes are related to each other by an affine transformation. At a point $p$, the coframe encodes an isomorphism between the cotangent space and the manifold. Such isomorphism is translated by the fact that the coframe can be used to rising or lowering the form rank of an object\footnote{For example, let $f^{(p)}$ be a fundamental $p$ form, then we can construct a $p-n$ form from $f^{(p)}$ as $f^{(p-n)}_{a_1a_2\ldots a_n}e^{a_1}e^{a_2}\ldots e^{a_n}$.}.\\\\
A direct consequence of the above described identification is that the premetric and metric can be identified through
\begin{eqnarray}
g_{\mu\nu}&\equiv&g_{ab}e^a\otimes e^b\;,\nonumber\\
g^{\mu\nu}&\equiv&g^{ab}e_a\otimes e_b\;,\nonumber\\
g^{ab}&=&e^a\cdot e^b\;,\nonumber\\
g_{ab}&=&e_a\cdot e_b\;,\label{metr01}
\end{eqnarray}
where we have assumed that the mapping $*\mathbb{T}^d_p\longleftrightarrow\mathbb{M}^d$, provided by the vielbein, preserves length measures. This is indeed the usual construction of a classical gauge theory for gravity \cite{Utiyama:1956sy,Kibble:1961ba,Mardones:1990qc,Zanelli:2005sa,Hehl:1994ue}. In the particular case of this work, the resulting class of theories is the metric-affine gravities \cite{Hehl:1976my,Hehl:1994ue}. Moreover, the relations \eqref{metr01}, through the spacetime metrics $g^{\mu\nu}$ and $g_{\mu\nu}$, imply that $g^{ab}$ is the inverse of $g_{ab}$ and \emph{vice versa}. Thus, such scenario provides a deep identification of the gauge space with the spacetime itself, giving rise to a geometrical theory, as gravity is supposed to be. Also, the premetric and its inverse can be employed to rise and low group indices and they provide a direct relation between group space and the dual group space. For example,
\begin{eqnarray}
e_a&=&g_{ab} e^b\;,\nonumber\\
e^a&=&g^{ab}e_b\;,\label{inv01}
\end{eqnarray}
which provides the relation between $T^a$ and $T_a$, namely
\begin{eqnarray}
T_a&=&g_{ab}T^b+Q_{ab}e^b\;,\nonumber\\
T^a&=&g^{ab}T_b+Q^{ab}e_b\;,\label{low01}
\end{eqnarray}
where we have used the relation $g^{ac}Q_{cb}=-g_{bc}Q^{ca}$. Moreover\footnote{Our
(anti-)symmetrization conventions are
\begin{eqnarray}
v_{(ab)}&=&v_{ab}+v_{ba}\;,\nonumber\\
v_{[ab]}&=&v_{ab}-v_{ba}\;.\nonumber
\end{eqnarray}}
\begin{equation}
Q_{ab}=\nabla
g_{ab}=\mathcal{D}g_{ab}=\mathrm{d}g_{ab}+\omega_{(ab)}\;,\label{nonmetr01}
\end{equation}
implying that, for a metric-affine gauge theory, even in a situation with constant metric tensor, $Q_{ab}\ne0$, due to the
symmetric sector of the connection. Thus, in metric-affine gravity, metric and connection are indeed incompatible quantities.\\\\
In a more sophisticated fashion, it is usually expected that the geometry is
generated by quantum effects of a well consistent quantum theory for
gravity. For instance, a metric tensor $g_{\mu\nu}$ would be
obtained from \cite{Mardones:1990qc,Zanelli:2005sa,Hehl:1994ue}
\begin{equation}
g_{\mu\nu}(x)=\left<g_{ab}(x)e^a_\mu(x)e^b_\nu(x)\right>\;,\label{metr02}
\end{equation}
which means that the expectation value of the composite operator
$g_{ab}(x)e^a_\mu(x)e^b_\nu(x)$ determines the space time
geometry. The expectation value \eqref{metr02} is supposed to be evaluated with a consistent quantum gravity path integral measure, which still lacks at the present human knowledge, see for instance \cite{Sobreiro:2007pn} and references therein.\\\\
Finally, let us clarify once more the subtle differences between the original metric-affine gravities \cite{Hehl:1976my,Hehl:1994ue} and our present interpretation. Firstly, in \cite{Hehl:1976my,Hehl:1994ue} vielbein and metric are both considered as potential fields with torsion and nonmetricity being their respective field strengths. In our case, as explained, those are considered matter fields. Moreover, $T$ and $Q$ do not arise from the covariant derivative algebra and $T$ equals itself to the translational curvature $\Xi$ only if translational gauge symmetry is broken, see \cite{Hehl:1994ue} and sect.\ref{Ee}. Secondly, affine gauge theory in curved spacetime and gravity as geometry are considered as non-overlapping sectors of the same theory. Such assumption is important in order to accommodate together the principles of the quantum theory and that of general relativity, see for instance \cite{Sobreiro:2007pn} and references therein.


\section{Fibre bundles and gauge theories}\label{FBT}

The formal mathematical description of gauge theories takes place on the
fibre bundle theory \cite{Kobayashi,Nakahara:1990th}. It is not our
aim to provide a review of fibre bundles and the respective
application to Physics. We attain ourselves to the specific case of
gauge theories, see for instance
\cite{Nakahara:1990th,Singer:1978dk,Daniel:1979ez,Bertlmann:1996xk}. For the present work, the fibre bundle theory will allow a formal characterization of topological and geometrical features of the scenario described in the previous section.

\subsection{Fibre bundle description of gauge theories}\label{ym}

Let us start with the simplest principal bundle for which a gauge
connection is generated \cite{Daniel:1979ez},
\begin{eqnarray}
{G(x)}&\equiv&\{G,\mathbb{M}^d,\pi,\phi\}\;,\nonumber\\
\pi&:&{G(x)}\longmapsto\mathbb{M}^d\;,\nonumber\\
\phi_i&:&{\pi}^{-1}(\{\mathbb{M}^d\}_i)\longmapsto\{\mathbb{M}^d\}_i\times
G\;,\label{ym01}
\end{eqnarray}
where the fibre and structure group are both a Lie group $G$, the
base space is a $d$-dimensional differential manifold $\mathbb{M}^d$,
usually the spacetime, and the total space is a nontrivial product
${G(x)}=G\circledast\mathbb{M}^d$. The projection map $\pi$ is a
continuous surjective map and the local triviality condition of the
total space is ensured by the homeomorphisms $\phi_i$, where
$\{\mathbb{M}^d\}_i$ are open neighbourhoods covering
$\mathbb{M}^d$. The definition of the principal bundle ${G(x)}$ is
simply a formal manner to describe the localization of the Lie group
$G$ in $d$-dimensional space, assembling to each point in
$\mathbb{M}^d$ a different value for the elements of $G$. The space
${G(x)}$ is then the local Lie group $G$.\\\\
The gauge connection rises on the definition of parallel transport
in the total space ${G(x)}$ \cite{Daniel:1979ez}. The gauge
transformations are associated with coordinates changing of the
total space with fixed base space coordinates,
\begin{eqnarray}
(x,r)&\longmapsto&(x,r^\prime)\;\big|\;x\in\mathbb{M}^d,\;\{r,r^\prime\}\subset G\;,\nonumber\\
r^\prime&=&g^{-1}r\;\big|\;g\in G\;.\label{map0}
\end{eqnarray}
In such transformation, which corresponds to a translation along the
fiber $\pi^{-1}(x)$, the connection transforms according to
\begin{equation}
A(x,r)\longmapsto A(x,r^\prime)=g^{-1}(x)\left(\mathrm{d}+A(x,r)\right)g(x)\;.\label{gt01}
\end{equation}
which is recognized as a typical gauge transformation for the gauge
potential, see \eqref{gauge02}.\\\\
Moreover, to every one-form connection $A$ there is a two-form
covariant curvature defined over $G(x)$, namely
\begin{equation}
F=\nabla^2=\mathrm{d}A+AA\;,\label{curv03}
\end{equation}
where
\begin{equation}
\nabla=\mathrm{d}+A\;.\label{covder04}
\end{equation}\\\\
The gauge connection is originated in ${G(x)}$ yet it does not
belong to the definition \eqref{ym01}. The complete description of
gauge theories follows from the product between ${G(x)}$ and the
space of all independent connections $A$, \emph{i.e.} those connections that are not related to each other by any gauge transformation. Thus, we may define
\cite{Singer:1978dk,Bertlmann:1996xk}
\begin{eqnarray}
\mathbb{Y}&\equiv&\{{G(x)},\mathcal{A},\pi,\phi\}\;,\nonumber\\
\pi&:&\mathbb{Y}\longmapsto\mathcal{A}\;,\nonumber\\
\phi_i&:&\pi^{-1}(\{\mathcal{A}\}_i)\longmapsto\{\mathcal{A}\}_i\times{G(x)}\;,\label{ym02}
\end{eqnarray}
where the fiber and structure group are both the local Lie group ${G(x)}$, the base
space $\mathcal{A}$ is the space of all independent algebra-valued
gauge connections $A$ and the total space is then a nontrivial
product $\mathbb{Y}={G(x)}\circledast\mathcal{A}$. A projection map
$\pi$ is assumed and the total space obeys local triviality
condition, which is characterized by the homeomorphisms $\phi_i$,
where $\{\mathcal{A}\}_i$ are open neighbourhoods covering
$\mathcal{A}$. The so called gauge orbit is obtained from
\eqref{gt01} by considering a field $A(x)\in\mathcal{A}$ and all
possible gauge transformations,
\begin{equation}
A^g=g^{-1}(\mathrm{d}+A)g\;,\label{ym03}
\end{equation}
which is exactly the fiber $\pi^{-1}(A(x))$. Thus, the total space
$\mathbb{Y}$ can be understood as the union of all gauge
orbits.\\\\
The main physical idea here is that ${G(x)}$ provides a local
character for $G$ while $\mathbb{Y}$ brings the connection to the
structure of the bundle. The principal bundle \eqref{ym01} provides
the existence of a gauge connection. However, this connection is
fixed, provided boundary conditions in $G(x)$. To give dynamics for the
connection one should consider all possible connections together
with a minimizing principle for a classical theory or a path
integral measure for a quantum one \cite{Daniel:1979ez,Bertlmann:1996xk}. This
dynamics is provided by the infinite dimensional principal bundle
\eqref{ym02}. In practice, $G(x)$ is more suitable for a system with
external interacting potential $A$ while $\mathbb{Y}$ for one which
the dynamics of $A$ is relevant.\\\\
Finally, let us remark that the present description is simply the
proper mathematical structure that describes a gauge theory over a
curved space. In fact, metric-affine gravity as described in Sect.2
fits this description since it is an affine gauge theory over a
curved manifold $\mathbb{M}^d$, assumed to be the spacetime.

\subsection{Reduction of principal bundles}\label{teo}

We shall now state three theorems from the fibre bundle theory that are of crucial importance in what follows. We shall not demonstrate them since they are known in the current literature that we shall proper refer to. The first one concerns the contraction of principal bundles, see \cite{Kobayashi,Nash:1983cq}.
\begin{itemize}
\item {\bf Theorem 1}: \emph{Let $\textbf{P}(G)$ denote a general principal
bundle with fibre characterized by a connected Lie group $G$, then
if $G=H\otimes C$,
where $H$ is a maximal compact subgroup of $G$ and $C$ is a
contractible space, then $G$ can be reduced to the subgroup $H$
resulting in the simpler principal bundle $\textbf{P}^\prime(H)$, as $\textbf{P}(G)\longrightarrow\textbf{P}^\prime(H)$. Thus, the principal bundle $\textbf{P}(G)$ is topologically
equivalent to the simpler principal bundle
$\textbf{P}^\prime(H)$.}\\\\
This theorem establishes that a principal bundle might be redundant, in the sense that it may be substituted by a smaller bundle with no loss of the topological information contained in it. It is obvious that the fundamental notion concerning this result is that of homotopy. The next theorem concerns connections on such principal bundles, see \cite{Kobayashi}, and it is just a necessary step for the third theorem.

\item {\bf Theorem 2}: \emph{Any connection on the reduced bundle $\textbf{P}^\prime(H)$ can be extended to a unique connection in $\textbf{P}(G)$.}\\\\
The contrary is however not always true, which reflects the fact that the space of connections on $\textbf{P}^\prime(H)$ does not cover the space of connections on $\textbf{P}(G)$. A connection in $\textbf{P}(G)$ that is obtained in the above way is said to be reducible to a connection in $\textbf{P}^\prime(H)$. An irreducible connection on $\textbf{P}(G)$ is such that it cannot be obtained by extending a connection on $\textbf{P}^\prime(H)$, see \cite{Kobayashi}. However, it might be possible in many cases that an irreducible connection on $\textbf{P}(G)$ characterize a connection on $\textbf{P}^\prime(H)$. The next theorem concerns whether such characterization is possible or not, see for instance \cite{Kobayashi,McInnes:1984kz}.

\item {\bf Theorem 3}: \emph{Let $A_G$ be the connection form on $\textbf{P}(G)$. Thus, the $H$ sector of the connection form, namely $A_H$, defines a connection in $\textbf{P}^\prime(H)$ if, and only if, the coset space $C=G/H$ is symmetric,} i.e. \emph{$[h,c]\subseteq c$, where $h$ and $c$ denote the Lie algebra of $H$ and $C$ sectors, respectively.}\\\\
This theorem is a direct consequence of the fact that, in symmetric spaces, the coset sector of the original connection responds covariantly to the action of an element of the subgroup $H$, which means that the coset sector behaves as a tensor space with respect to $H$.
\end{itemize}
Our next task is to understand the consequences of the above theorems on metric-affine gauge theories as constructed in Sect.\ref{MAG}.


\section{Fibre bundles and metric-affine gravities}\label{FBMAG}

\subsection{Affine group bundle and its contraction}

Following the description of Sect.\ref{ym}, an affine algebra-valued connection is generated from the principal bundle \eqref{ym01} by assuming $G=A_{\mathrm{rigid}}(d,\mathbb{R})$, where $A_{\mathrm{rigid}}(d,\mathbb{R})$ is the global affine group, and the base space a manifold with $d$ dimensions, $\mathbb{M}^d$. Let us denote it by
\begin{equation}
A(d,\mathbb{R})\equiv\left\{A_{\mathrm{rigid}}(d,\mathbb{R}),\mathbb{M}^d,\pi,\phi\right\}\;,\label{a1}
\end{equation}
that corresponds to the localization of the affine group. For precision purposes we name the bundle \eqref{a1} \emph{affine gauge bundle}. Moreover, the fact that the base space dimension coincides with the Casimir $d$ of the gauge group ensures the possibility of identifying the group space with the cotangent bundle, as required by gravity.\\\\
The connection that rises from \eqref{a1} is obviously the $A(d,\mathbb{R})$ connection $Y$ while the corresponding curvature is $\Theta$, both defined in Sect.\ref{gauge00}. The matter fields $e^a$, $e_a$, $g^{ab}$ and $g_{ab}$ are introduced as tensor representations of the group space. As required $\overline{Y}$ is also introduced.\\\\
The application of Theorem 1 to the principal bundle
\eqref{a1} is quite direct and it can be
done in two steps. First, by collapsing the translational sector, we
reduce the affine group to the general linear group, according to
the properties discussed in Sect.\ref{grstr}. This task is easily achieved
since $\mathbb{R}^d$ is trivially contractible
\cite{Kobayashi,Nash:1983cq}. The second step is to reduce the general
linear group to a maximal compact subgroup, in this case we chose
the orthogonal group\footnote{There are several other possibilities of
contraction, for instance $O(d-p,p)$, $SL(d,\mathbb{R})$,
$Sp(d,\mathbb{R})$ or $GL(d/2,\mathbb{C})$. However, the orthogonal
group is the only case that there are no extra necessary conditions
for the contraction. For example, a Lorentzian contraction
$O(d-1,1)$ requires that $\mathbb{M}^d$ would have vanishing Euler
characteristic. Also, for noncompact manifolds, the reduction to
Lorentzian structures are always possible. See \cite{Nash:1983cq}.}. This
is also doable since the coset space $K(d)$
is also trivial\footnote{The elements of $K(d)$ are parameterized as
$e^{b^a_{\phantom{a}b}\lambda^b_{\phantom{b}a}}$ where $\lambda$ are
symmetric matrices. This parametrization defines a map
$f:S(d)\mapsto K(d)$, where $S(d)$ is the space of all real
invertible $d\times d$ symmetric matrices. The space of symmetric
matrices is a vector space and thus homeomorphic to $\mathbb{R}^d$,
which is contractible. See \cite{Nash:1983cq}.}. The contraction of the affine group is then
a continuous deformation,
\begin{equation}
A_{\mathrm{rigid}}(d,\mathbb{R})\longrightarrow GL_{\mathrm{rigid}}(d,\mathbb{R})\longrightarrow O_{\mathrm{rigid}}(d)\;,\label{contra01}
\end{equation}
which implies that
\begin{equation}
A(d,\mathbb{R})\longrightarrow GL(d,\mathbb{R})\longrightarrow O(d)\;.\label{contra02}
\end{equation}
where the orthogonal bundle is defined as
\begin{equation}
O(d)=\left\{O_{\mathrm{rigid}}(d),\mathbb{M}^d,\pi,\phi\right\}\;.\label{o1}
\end{equation}
Now, since $\mathbb{R}^d$
and $K(d)$ are symmetric, see \eqref{alg000} and
\eqref{coset0}, we can apply Theorem 3.
The full connection $Y$ can be indeed restricted to the orthogonal
sector $w$. As discussed in \cite{McInnes:1984kz}, the main result
here is that every connection on $A(d,\mathbb{R})$ imposes a
connection on $O(d)$.

\subsection{Vielbein-coframe identificantion and second order gravity}

The relevance for gravity is visualized by taking the bundle \eqref{a1} and identifying it with the coaffine bundle \cite{Trautman:1979cq,Trautman:1981fd,Hennig:1981id}  throught the vielbein-coframe identification. Under this scenario the first contraction in \eqref{contra01} implies on the topological equivalence between the coaffine and coframe bundles while the second one shows that both bundles are equivalent to the orthogonal coframe bundle \eqref{o1}.  The reduction of the metric-affine geometry to the Riemann-Cartan one is then established. In fact, the premetric now transforms only under the orthogonal group. Thus, it suffers only rigid rotations characterizing a Riemannian metric \cite{Nash:1983cq,McInnes:1984kz}. To see this one can take for instance \eqref{gtransfinf} for the case of orthogonal transformations. This restriction results on the invariance of $g$ under orthogonal transformations, $g^{ab}\longmapsto g^{ab}$, implying that $g^{ab}$ is of the same equivalence class of a flat metric, $g^{ab}\sim \delta^{ab}$ and thus $Q^{ab}\sim 0$. From Theorem 3 the connection can be directly restricted to the orthogonal sector \cite{McInnes:1984kz}, $Y\rightarrow w$, resulting on the usual Riemann curvature $\Omega\rightarrow\Omega(w)$ and the usual Cartan torsion $T\rightarrow T(w,e)$.\\\\
In \cite{McInnes:1984kz} a similar contraction exercise was performed for the Poincar\'e group, $A(d,\mathbb{R})\longrightarrow
O(d)\ltimes\mathbb{R}^d=ISO(d)$. In this case, the the coset
$K^\prime(d)=A(d,\mathbb{R})/ISO(d)$ is not a symmetric space since
$[k^\prime,iso]\subseteq k^\prime\oplus iso$. Thus, a connection
$Y\in A(d,\mathbb{R})$ does not impose a connection for a Poincar\'e
gauge theory. In this case the compatibility condition cannot be
derived from the contraction of the coaffine bundle to the
Poincar\'e one unless $Q=0$ is imposed by hand.\\\\
It is clear that the coaffine, coframe or orthogonal coframe bundles are proper scenarios for a second order gravity theory, \emph{i.e.}, a theory for a dynamical vielbein and non-dynamical connection. Obviously, the metric tensor is no longer independent due to the identification \eqref{metr01}. In this case, it could be possible, as it is in gravity, to obtain the connection as a function of the dynamical quantities, $w(e)$. Remarkably, in the bundle $O(d)$, this would lead to a pure Riemannian connection implying that $T=0$. The reason is simple, the unique dynamical field is $e$ and to provide nonvanishing torsion one should introduce a source for it, for example, spinor matter fields. This fits with general relativity where there is only one dynamical field, $e$, and all physical matter are classical distributions in spacetime. Moreover, in general relativity, any point particle is supposed to travel along geodesics whose equation does not depend on torsion and nonmetricity, implying that point particles would never be able to feel such geometric features. To construct a theory that the connection is indeed a dynamical field, just like a gauge theory is, one should consider a more sophisticated bundle just like \eqref{ym02}.

\subsection{Dynamical affine gauge bundle and its contraction}

To consider a more fundamental gravity theory we need to provide dynamics to the connection $Y$. To do so, we consider the principal bundle \eqref{ym02} with
$G(x)=A(d,\mathbb{R})$ and base space given by the space of all
independent connections $\mathcal{Y}$ that can be defined in
\eqref{a1},
\begin{equation}
\mathbb{A}\equiv\{A(d,\mathbb{R}),\mathcal{Y},\pi,\phi\}\;.\label{a2}
\end{equation}
The base space $\mathcal{Y}=\mathbb{Y}/A(d,\mathbb{R})$, also called moduli space, can be decomposed
into the spaces of all $GL(d,\mathbb{R})$ connections $\mathcal{W}$
and all translational connections $\mathcal{E}$, namely
$\mathcal{Y}=\mathcal{W}\oplus\mathcal{E}$. A typical fibre
$\pi^{-1}(Y(x))$, where $Y(x)\in\mathcal{Y}$, is given by a gauge
orbit
\begin{equation}
\pi^{-1}(Y(x))\;:\;Y^{\mathcal{U}}=\mathcal{U}^{-1}\left(\mathrm{d}+Y\right)\mathcal{U}\;,\label{orb01}
\end{equation}
characterizing gauge transformations. To avoid confusion with the simpler bundle \eqref{a1} we call the bundle \eqref{a2} by \emph{dynamical affine gauge bundle}.\\\\
For the principal bundle \eqref{a2} we confine ourselves to Theorem 1. The reason is simple, the induced connection on $\mathbb{A}$ is not that of \eqref{a1} \cite{Bertlmann:1996xk}. In fact, the connection that is generated in $\mathbb{A}$ is not relevant for gauge theories at the present level. The application of Theorem 1 to $\mathbb{A}$ can be performed by first applying it to its structure group $A(d,\mathbb{R})$ (previous section) and then, by taking into account that all gauge connections and their equivalents were brought into the structure
of the bundle \eqref{a2} before the contraction, analyze the respective effect on a typical fibre \eqref{orb01}. To do so, we decompose the full affine connection according to
\begin{equation}
Y=w+q+E\;,\label{decomp01}
\end{equation}
where $w$ is the orthogonal sector of the connection while $q$ the
$K(d)$ one,
\begin{eqnarray}
w&=&\omega^a_{\phantom{a}b}\Sigma^b_{\phantom{b}a}\;,\nonumber\\
q&=&\omega^a_{\phantom{a}b}\lambda^b_{\phantom{b}a}\;,\label{decomp02}
\end{eqnarray}
Let us take then a generic fibre \eqref{orb01} and contract the group space according to $\mathcal{U}\longrightarrow L$. Thanks to the symmetric nature of the coset spaces $\mathbb{R}^d$ and $K(d)$, the resulting decomposition is
\begin{eqnarray}
w^L&=&L^{-1}(\mathrm{d}+w)L\;,\nonumber\\
q^L&=&L^{-1}qL\;,\nonumber\\
E^L&=&L^{-1}E\;,\label{decomp03}
\end{eqnarray}
\emph{i.e.}, there are no mixing between the coset sectors and the orthogonal one. Moreover, for the matter sector, the contraction
implies in
\begin{eqnarray}
e^L&=&L^{-1}e\;,\nonumber\\
g^L&=&L^{-1}g\left(L^{-1}\right)^T\;.\label{decomp04}
\end{eqnarray}
Obviously, analogous relations for the dual fields are also obtained. We may conclude that the non-orthogonal sector of $Y$, after the
contraction \eqref{contra02}, abandon the geometric sector to migrate to the matter sector. Thus,
the metric-affine dynamical geometry described by \eqref{a2}
contracts down to a Riemann-Cartan one while the non-orthogonal
components of the original connection are thrown to the matter
sector. This can be described by
\begin{equation}
\mathbb{A}\longrightarrow\mathbb{O}\oplus\mathcal{E}\oplus\mathcal{Q}\;,\label{contra03}
\end{equation}
where
\begin{equation}
\mathbb{O}\equiv\{O(d),\mathcal{L},\pi,\phi_i\}\;,\label{o2}
\end{equation}
is the dynamical principal bundle for a Riemann-Cartan theory. The
space $\mathcal{L}\subset\mathcal{W}$ is the space of all independent orthogonal
connections and $\mathcal{Q}=\mathcal{W}/\mathcal{L}$ is the space of the $K(d)$ connections. We can say
that \emph{If one starts with a dynamical metric-affine gauge theory
of gravity it can end up in a Riemann-Cartan theory with two extra
matter fields.} In fact, the curvature decomposition yields
\begin{eqnarray}
\Omega^a_{\phantom{a}b}&=&R^a_{\phantom{a}b}+V^a_{\phantom{a}b}\;,\nonumber\\
\Xi^a&=&\mathrm{D}E^a-q^a_{\phantom{b}b}E^b\;,
\label{decompcurv}
\end{eqnarray}
with
\begin{eqnarray}
R^a_{\phantom{a}b}&=&\mathrm{d}w^a_{\phantom{a}b}-w^a_{\phantom{a}c}w^c_{\phantom{c}b}\;,\nonumber\\
V^a_{\phantom{a}b}&=&\mathrm{D}q^a_{\phantom{a}b}-q^a_{\phantom{a}c}q^c_{\phantom{b}b}\;.\label{curvO}
\end{eqnarray}
where $D=\mathcal{D}(w)$ is the orthogonal covariant derivative. Obviously, $R$ is the orthogonal field strength while $V$ is the orthogonal minimal coupling of $q$ with a nonlinear mater interaction term. The same interpretation follows for $E$ and its coupling $\Xi$. For the minimal coupling of the vielbein we achieve also the orthogonal minimal coupling for $e$ and a matter interaction term,
\begin{equation}
T^a=\mathrm{D}e^a-q^a_{\phantom{a}b}e^b\;.\label{torx}
\end{equation}
Finally, for $Q$ it is easy to show that the same behavior is encountered,
\begin{equation}
Q^{ab}=\mathrm{D}g^{ab}-q^a_{\phantom{a}c}g^{cb}-q^b_{\phantom{b}c}g^{ca}\;,\label{mincoup01}
\end{equation}
Obviously, analogous relations are obtained for all dual fields.

\subsection{Identifying vielbein and coframes}

We are now dealing with the principal bundle \eqref{o2}. Moreover, the nonorthogonal fields do not decouple that easy since they carry dynamics through the original dynamical gauge bundle \eqref{a2}. After identifying coframes and vielbein we have to keep in mind that the group space is now flat with a deformed rigid dynamical premetric. We can write then
\begin{eqnarray}
g^{ab}&=&\delta^{ab}+\gamma^{ab}\;,\nonumber\\
g_{ab}&=&\delta_{ab}+\gamma_{ab}\;,\label{exp0}
\end{eqnarray}
in such a way that $\gamma^{ab}$, and its dual, corresponds to the post-Riemannian sector of the original metric tensor \eqref{metr01}. Thus, $\delta^{ab}$ will work as the flat metric of cotangent space while $\gamma^{ab}$ will be a generic covariant matter field, symmetric in the group indices. In fact, from \eqref{gtransf}, we obtain
\begin{eqnarray}
\delta&\longmapsto&\delta\;,\nonumber\\
\gamma&\longmapsto&L^{-1}\gamma\left(L^{-1}\right)^T\;.\label{gammatransf}
\end{eqnarray}
The identification is then performed as
\begin{eqnarray}
g_{\mu\nu}&=&\delta_{ab}e^a\otimes e^b\;,\nonumber\\
g^{\mu\nu}&=&\delta^{ab}e_a\otimes e_b\;,\label{idx}
\end{eqnarray}
ensuring the Riemannian character of $\mathbb{M}^d$. The affine connection decomposes as \eqref{decomp01}, \emph{i.e.}, into an orthogonal connection $w^a_{\phantom{a}b}$, a symmetric matter field $q^a_{\phantom{a}b}$ and a translational matter field $E^a$. Their transformations are now described by \eqref{decomp03}. As a consequence of \eqref{decomp01} and \eqref{idx}, expression \eqref{torx} reduces to
\begin{equation}
T^a=\tau^a=\mathrm{D}e^a\;,\label{torxx}
\end{equation}
while \eqref{mincoup01} reduces to
\begin{equation}
Q^{ab}=\mathrm{D}\gamma^{ab}-2q^{ab}-q^a_{\phantom{a}c}\gamma^{cb}-q^b_{\phantom{b}c}\gamma^{ca}\;,\label{mincoup02}
\end{equation}
Expression \eqref{torxx} is the usual Cartan torsion \cite{Kibble:1961ba,Mardones:1990qc,Zanelli:2005sa} for the orthogonal group while \eqref{mincoup02} actually is a covariant quantity under orthogonal gauge transformations and carries no geometrical information. The matter nature of $\gamma$ is clear from its minimal coupling with $w$ in \eqref{mincoup02} and from its transformation law \eqref{gammatransf}. From \eqref{curvO} it is evident the matter character of $q$ while the usual Riemann curvature of the spacetime is associated with the field strength of the reduced orthogonal gauge theory. It turns out that $\gamma$ and $q$ are now just a kind of memory of the original gauge symmetry not related with the spacetime geometry. The same understanding holds for the object $Q^{ab}$, \emph{i.e.}, it does not concern nonmetricity. In fact, a direct computation shows that nonmetricity vanishes on the reduced bundle \eqref{o2} since the metric tensor is now originated from a flat cotangent space. It is worth mention that expression \eqref{exp0} has nothing to do with a background expansion, first because it is based on consistent geometrical features of the gauge structure of the theory and second because $\gamma$ does not need to be small when compared to $\delta$.\\\\
Another important consequence is that now one can identify $E$ and $e$ without the necessity of introducing the compensating field $\varphi$ since they have now the same transformation rule, see \eqref{decomp03} and \eqref{decomp04}. As a consequence $\Xi=\tau$, as can be deduced from \eqref{decompcurv} and \eqref{torxx}.\\\\
Some remarks are now in order. First we recall that the same results are obtained by performing the identification before the contraction, according to expression \eqref{metr01}. A \emph{posteriori} decomposition of the kind of \eqref{exp0} can be performed and the post-Riemannian sector of the spacetime metric destabilizes since it does not suffer any other deformation induced by gauge transformations, now pure orthogonal transformations. Let us also remark that, If we have started directly from the orthogonal bundle \eqref{o1} as the fundamental structure for gravity \cite{Kibble:1961ba,Mardones:1990qc,Zanelli:2005sa}, we would have $E=\gamma=q=0$, already from the beginning, which is the usual way to construct gravity in the first order formalism for the orthogonal gauge group. Finally, we point out that the contractions discussed in this section are not enforced by dynamical effects such spontaneous or dynamical symmetry breaking, they follow simply by the topological nature of the gauge theory. In this sense, the contraction of the dynamical bundle \eqref{a2} are optional, which means that one can simply choose to work in the original bundle \eqref{a2} where all degrees of freedom have geometrical interpretation through \eqref{metr01} or in the contracted one \eqref{o2} and deal with additional matter fields but with a very much simpler geometry by using \eqref{idx}.


\section{Conclusions}\label{FINAL}

In this work we have proposed a suitable scenario for the construction of a quantum theory of metric-affine gravities by stating the independence between the internal gauge space and the cotangent bundle of spacetime. This independence is ensured by considering vielbein, premetric and their duals as genuine matter fields. Although we have not provide a specific consistent quantum theory, we have studied some geometrical properties of the formalism. Further, the present framework allow for the compatibility of the principles of general relativity and those of quantum physics by assuming that both sectors shall not overlap, \emph{i.e.}, a consistent quantum model should be able to maintain the gauge internal space independent of the spacetime. Thus, dynamical effects are supposed to generate the identification between the cotangent space and gauge space, for instance by the vielbein-coframe identification \eqref{metr02}. Thus, the theory is automatically thrown in a geometric classical sector. We recall that this idea was already discussed for a simpler model in \cite{Sobreiro:2007pn} under the Palatini formalism.\\\\
The main result of this article is the reduction of the metric-affine geometry to the Riemann-Cartan one. By considering the correct fibre bundle description of a gauge theory for the affine group \eqref{a2}, we have consistently contracted it to the orthogonal bundle \eqref{o2}. Thus, by decomposing the relevant fields into their orthogonal sector and the rest it is shown that we achieve a gauge theory for the orthogonal gauge group and extra matter fields besides vielbein and premetric. In fact, the translational connection and the symmetric sector of the $GL(d,\mathbb{R})$ connection migrate to the matter sector.\\\\
After the reduction the theory is then mapped into a geometric theory as required by gravity. This is performed by identifying the vielbein matter field with coframes belonging to the cotangent bundle. Moreover, by decomposing the premetric into a flat sector and a post-Riemannian one, we can then map the flat sector into the spacetime metric by means of \eqref{idx} and thus end up in a Riemann-Cartan gravity with extra sectors of matter. The original setup is composed by the gauge fields $\omega$ and $E$ and the matter fields $e$, $g$ and the respective duals charactering the quantum sector. The reduced mapped sector is characterized by the orthogonal gauge field $w$, the vielbein field $e$ (coframes) and the matter fields $q$, $E$ and $\gamma$. The last is supposed to describe the classical gravity as a geometric theory.\\\\
We recall that the construction of a general invariant action for the metric-affine gravity and the respective reduction will be part of a different paper due to its extension and intricacies involved. We refer to \cite{Hehl:1999sb} for an extensive analysis for constructing metric-affine gauge invariant actions and the respective possible solutions by assuming already the gauge space association with spacetime. However, we can mention at least one example of such reduction \cite{Sotiriou:2009xt}. In this example, a metric-affine $f(R)$ gravity is shown to be reduced to a $\omega_0=-3/2$ Brans-Dicke theory. Another point to be investigated in the future is the viability on the quantization of the proposed scenario, a highly nontrivial task. Moreover, the effect of the reduction on additional matter fields is also left to a future work as well as the possibility of the new matter fields $E$, $q$ and $\gamma$ to be related to dark matter/energy. In particular, the introduction of fermionic matter fields should be done via the so called multispinors
\cite{Ne'eman:1977du,Ne'eman:1978gj} and the respective consequences for the geometry contraction is quite obscure at the present situation. In summary, in this work we provide a few ideas that goes on the direction of a possible quantum framework for a gauge theory of gravity as well as the formal analysis for the reduction of metric-affine geometry to the Riemann-Cartan one, independently of the starting action, resulting on the migration of the original nonmetric degrees to the matter set of fields.


\section*{Acknowledgements}

We are thankful to the referee for bring to our attention references \cite{Hennig:1981id,Lord:1986xi,Lord:1986at,Lord:1988nd}. The authors express their gratitude to the Conselho Nacional de Desenvolvimento Cient\'{i}fico e Tecnol\'{o}gico\footnote{RFS is a level PQ-2 researcher under the program \emph{Produtividade em Pesquisa}, 304924/2009-1.} (CNPq-Brazil) for financial support. RFS is also partially supported, and gratefull for, by the Funda\c{c}\~ao Carlos Chagas Filho de Amparo \`a Pesquisa do Estado do Rio de Janeiro\footnote{Under the program \emph{Aux\'\i lio Instala\c{c}\~ao}, E-26/110.993/2009.} (FAPERJ) and the Pro-Reitoria de Pesquisa, P\'os-Gradua\c{c}\~ao e Inova\c{c}\~ao\footnote{Under the program \emph{Jovens Pesquisadores 2009}, project 304.} of the Universidade Federal Fluminense (Proppi-UFF).


\end{document}